\documentclass{elsart}

\usepackage{psfig}

\usepackage{natbib}

\begin{document}

\runauthor{A. Comastri}


\begin{frontmatter}

\title{The BeppoSAX view of NLS1s}

\author[OAB]{Andrea Comastri}
\address[OAB]{Osservatorio Astronomico di Bologna, via Ranzani 1, I-40127 
Bologna, Italy}

\begin{abstract}
The main results of broad band (0.1--10 keV) BeppoSAX observations of a 
selected sample of NLS1s are presented and discussed. It is shown that 
all the available data are consistent with a scenario in which NLS1s are
running at a high accretion rate.
\end{abstract}


\end{frontmatter}


\section{Introduction}

An observational program of a small sample of Narrow Line Seyfert 1 
Galaxies (NLS1s) has been carried out with BeppoSAX with the aim to 
investigate the broad band X--ray spectral and variability properties 
of these objects. 
The capabilities of the BeppoSAX detectors, 
and especially the relatively large MECS effective area at 
high energy ($>$ 5 keV), have been fully exploited to further 
investigate, with respect to previous ROSAT and ASCA observations, 
several of the distinctive properties of NLS1s.
More specifically, we want to test whether the 2--10 keV spectral
index 
distribution and the properties of the iron K--shell features
in the 6--10 keV region are different from those of normal, 
broad--line Seyfert 1s (BLS1s).
In addition, the broad energy range covered by the BeppoSAX LECS
(0.1--4 keV),  
MECS (2--10 keV), and PDS (13--100 keV) detectors will be used to 
constrain the overall shape of the X--ray continuum and in particular
the strength of the soft excess component and the nature of the 1 keV
absorption/emission features reported in several ASCA 
observations \cite{L97}, \cite{F98}, \cite {TGN98}, \cite{L992},
\cite{TGN99}, \cite{V991}, \cite{V992}.
In this paper, a summary of the most 
important results obtained by BeppoSAX are presented and discussed.
A more detailed analysis of the X--ray observations 
complemented by optical and UV data for some sources can be found
elsewhere : \cite{C98}, \cite{C00}, \cite{P00}.  

\section{The sample}

The BeppoSAX NLS1 Core Program includes some of the brightest and most 
variable objects previously observed by ROSAT and/or ASCA (Table 1). 
We have also considered the optically selected
quasar PG~1115+407, observed in a different BeppoSAX program, since its optical
and X--ray properties satisfy the NLS1 definition. 

All datasets were analyzed in a uniform way 
using the standard data reduction techniques described in \cite {FGG99}.
All objects have been clearly detected by the imaging 
LECS and MECS detectors, while only two positive detections
(PKS~0558$-$504 and PG~1115+407) are found in the PDS instrument.
Unfortunately, in both cases the spectrum above 10 keV is poorly
constrained. For PKS~0558$-$504, the PDS data are consistent with the
extrapolation of the 2--10 keV spectrum.  A spectral flattening seems to be 
present in PG~1115+407 which may be  due to contaminating
sources in the field of view \cite{M00}.

\begin{table*}
\centering
\caption{The BeppoSAX sample}
\vspace{0.05in}
\begin{tabular}{lccccc}
\hline
~ & ~ & ~ & ~ & ~ & ~  \\
Source & Date & \multicolumn{2}{c}{Exposures (sec)} &
\multicolumn{2}{c}{Count Rate (cts/sec)}  \\
~ & ~ & LECS  & MECS &  LECS$^a$  & MECS$^b$   \\
\hline
Ark~564 & 14/11/97 & 12574 & 26196 & 0.254$\pm$0.005 & 0.170$\pm$0.003  \\
Ark~564 & 12/06/98 & 20209 & 46765 & 0.250$\pm$0.004 & 0.169$\pm$0.002  \\
Ark~564 & 22/11/98 & 11963 & 29154 & 0.255$\pm$0.005 & 0.186$\pm$0.003  \\
\hline
PKS~0558$-$504 & 18/10/98 & 33169 & 64021 & 0.126$\pm$0.002 & 0.165$\pm$0.002 \\
\hline
Ton~S~180 & 03/12/96 & 12014 & 24913 & 0.092$\pm$0.003 & 0.068$\pm$0.002 \\
\hline
IRAS~13349+2438 & 13/01/00 & 25327 & 69552 & 0.056$\pm$0.002 & 
0.053$\pm$0.001 \\
\hline
RE~J1034+396 & 18/04/97 & 21347 & 43183 & 0.066$\pm$0.002 & 0.016$\pm$0.001 \\
\hline
PG~1115+407 & 02/05/97 & 17078 & 33798 & 0.016$\pm$0.001 &  0.012$\pm$0.001 \\ 
\hline
IRAS~13224$-$3809 & 29/01/98 & 17304 & 39279 & 0.009$\pm$0.001 &
0.006$\pm$0.0006 \\
\hline

\end{tabular}
\par\noindent
$^a$ in the 0.2--2 keV energy range; $^b$ in the 2--10 keV energy range

\end{table*}

\section{The 2--10 keV spectrum}

The MECS datasets of the entire sample were fitted with 
a single power law model plus Galactic absorption.
This model provides an acceptable description of the X--ray continuum
and the results of individual fits are reported in Table 2.
The average slope is $\alpha$= 1.13 with a large associated dispersion 
of $\sigma$=0.31. Excluding the faintest source in the sample
(IRAS~13224$-$3809), 
for which the slope is poorly constrained, we find  
$ \langle \alpha \rangle$ = 1.22$\pm$ 0.23, fully consistent with the 
weighted average value of 1.23 for the whole sample. 
The small number of objects does not allow a 
detailed comparison with the 2--10 keV slopes  
of BLS1s and quasars, even though it is clear that the 
2--10 keV energy index distribution of NLS1s is shifted 
toward higher values compared to that of BLS1s, in agreement with 
previous findings \cite{BME97}, \cite{L992}, \cite{V992}.
A steep 2--10 keV spectrum is expected by two--phase thermal 
Comptonization models \cite{HM93} if a significant 
fraction of the accretion power is dissipated 
in the disc phase. This hypothesis is supported by the presence
of strong soft components in most of the objects (see $\S$ 6).

\begin{table*}
\centering
\caption{MECS spectral fits in the 2--10 keV energy range}
\vspace{0.05in}
\begin{tabular}{lccccc}
\hline
 ~ & ~ & ~ & ~ & ~ & ~ \\
Source & $z$ & $\alpha$ & $\chi^2$/d.o.f. & $F^a_{2-10 keV}$ & $L^b_{2-10 keV}$   \\
\hline
Ark~564$^c$ & 0.025  &  1.41$\pm$0.03 & 95.2/68 & 14.7 & 0.4 \\
\hline
PKS~0558$-$504 & 0.137 & 0.97$\pm$0.04 & 134.5/141 & 15.4 & 14 \\
\hline
Ton~S~180 & 0.062 &  1.28$\pm$0.12  &  69.1/64  & 3.9  & 0.7 \\
\hline
IRAS~13349+2438 & 0.108  & 0.89$\pm$0.09 & 104.6/96 & 3.6 & 2.0  \\
\hline
RE~J1034+396 & 0.042 &  1.42$\pm$0.26 & 31.5/45 & 0.86 & 0.07 \\
\hline
PG~1115+407 & 0.154 & 1.35$\pm$0.27 & 23.0/26 & 0.85 & 1.1 \\
\hline
IRAS~13224$-$3809 & 0.067  & 0.62$\pm$0.45 & 16.6/17 & 0.49 & 0.1  \\
\hline
\end{tabular}
\par\noindent
$^a$ units of 10$^{-12}$ erg cm$^{-2}$ s$^{-1}$ 
\par\noindent
$^b$ units of 10$^{44}$ erg s$^{-1}$ ; 
$H_0$ = 50 km s$^{-1}$ Mpc$^{-1}$ , $q_0$=0 
\par\noindent
$^c$ 3 observations merged
\end{table*}

\section{Hard X--ray variability}

Substantial flux variability (up to a factor of 2) in both the soft and hard
X--ray bands is detected in all sources except REJ~1034+396
on timescales of the order of a few thousands of seconds, 
confirming the temporal behaviour established by ROSAT 
\cite{BBF96} and ASCA \cite{L991}, \cite{T99} observations. 
The analysis of hardness--ratio light curves indicates the lack of
significant spectral variability \cite{C98}, \cite{C00}.
After the discovery of extremely rapid (doubling time of about 800 s) 
soft X--ray variability in the ROSAT observation of IRAS~13224$-$3809
\cite{B93}, several intensive monitoring programs of a few NLS1s 
have been successfully carried out with the ROSAT HRI, allowing 
a better sampling of the light curves.
Several episodes of rapid soft X--ray variability with extremely high amplitude 
(up to a factor of 50--60) have been detected 
in IRAS~13224$-$3809 \cite{B97}, PHL 1092  \cite{B99} and PKS~0558$-$504
\cite{G00}. 
Relativistic effects and/or obscuration by thick matter with a small covering
factor have been proposed (see Brandt, this volume).

The BeppoSAX 2--10 keV light curve of the rapidly variable 
NLS1 IRAS 13224$-$3809 (Fig.~1) indicates that relatively rapid high amplitude 
variability is also present at high energies. 
Unfortunately the faint X--ray state and the short exposure time 
do not allow a more detailed analysis and a comparison between soft and 
hard X--ray light curves.

\begin{figure}[htb]
\centerline{\psfig{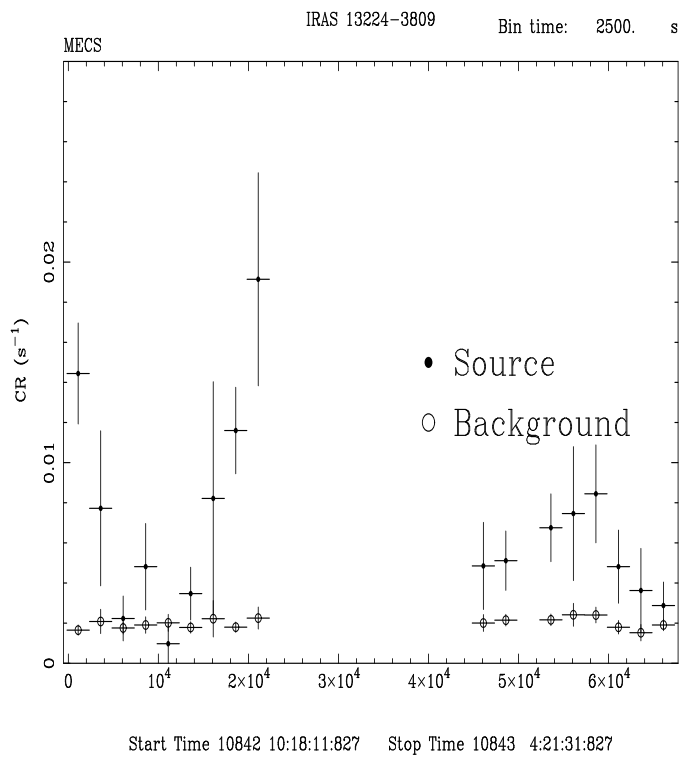}}
\caption{The hard X--ray light curve of IRAS~13224$-$3809.}
\end{figure}

\section{Iron Features}

The search for iron K$\alpha$ emission lines gave positive results,
although at different levels of significance, in 
3 objects: Ton~S~180 \cite{C98}, Ark~564 \cite{C00}
and PG~1115+407 \cite{M00}.
The best fit energy centroid, significantly
higher than 6.4 keV (between 6.7 and 7 keV), and the observed 
equivalent widths of the order of 500 eV in Ton~S~180 and PG~1115+407, 
strongly suggest emission from highly ionized gas.
There is no evidence of line--like features in the other objects 
but it should be noted that the signal to noise at high energies is rather 
poor. The lack of line emission in the good quality spectrum of 
the radio--loud quasar PKS~0558$-$504 can be explained assuming a 
substantial contribution from the non--thermal component.  

\begin{figure}[htb]
\centerline{\psfig{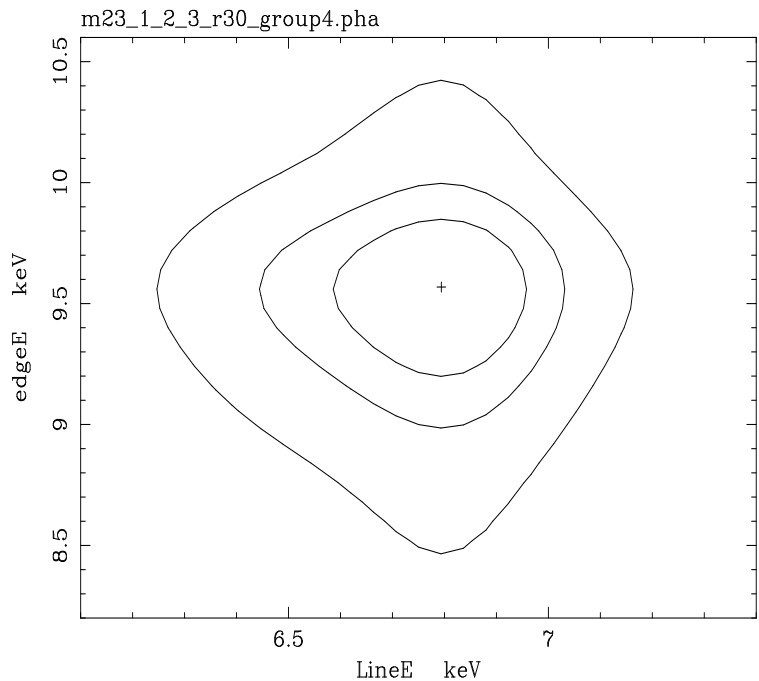}}
\caption{The 68, 90 and 99\% confidence contours of the iron line and edge
energies in the BeppoSAX spectrum of Ark~564.}
\end{figure} 

The residuals of a single power law fit to the high energy spectrum 
of Ark~564 suggest the presence of an edge--like feature at E$>$ 8 keV
in addition to the ionized iron line (Fig.~2).
The best fit energy and optical depth of the absorption edge
are consistent with those recently reported from the analysis of 
simultaneous ASCA and RXTE observations \cite{V991}.

The presence of emission lines and absorption edges 
originating in a highly ionized environment seems to be a distinctive 
property of the BeppoSAX NLS1 sample. 
In this respect it is worth noting the tentative detection 
(2$\sigma$) of a line--like excess at about 6.7 keV in the ASCA spectrum 
of RE~J1034+396 \cite{P95}. A straightforward interpretation of the 
line properties is obtained if they are produced in the 
surface layers of a strongly ionized accretion disc \cite{M93}.

\section{Broad band spectra}

The LECS plus MECS data of all sources have been fitted with 
various models leaving the relative normalizations free to vary 
in order to take into account residual intercalibration systematic 
uncertainties \cite{FGG99}.
A single component model never provides an acceptable fit
to the 0.1--10 keV continuum. The residuals of a single power law
fit to the IRAS~13349$+$2438 spectrum (Fig.~3) are
representative of the typical behaviour observed in the
BeppoSAX NLS1 sample and clearly illustrate the need for at least two
components.

\begin{figure}[htb]
\centerline{\psfig{figure=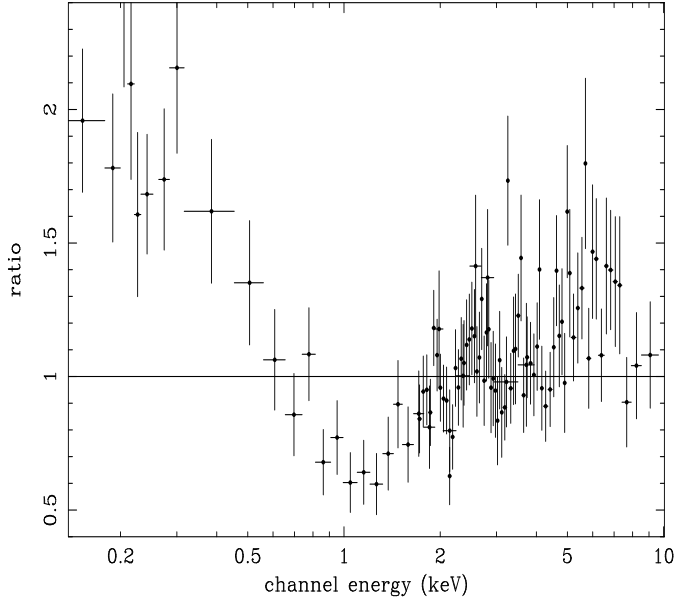,height=3.5truein,width=3.9truein,angle=-90}}
\caption{The residuals of a single power law fit to the 0.1--10 keV
spectrum of IRAS~13349$+$2438.}
\end{figure}

The best fit spectral parameters, assuming Galactic absorption 
and neglecting the high energy emission/absorption features,
are summarized in Table 3 together with the corresponding unabsorbed 
luminosities in the 0.1--2 and 2--10 keV bands. Even though the soft 
X--ray flux is always larger than that in the hard X--rays, there is a 
relatively wide range in the shape and intensity of the soft excess component.
The high energy tail of blackbody emission provides a good description 
of the soft X--ray spectrum for 5 (out of 7) objects.
Not surprisingly, the strongest soft excesses (last column of table 3) 
correspond to the highest temperatures and an additional 
blackbody component is required to fit the lowest energy 
(below 0.3--0.4 keV) part of the spectrum in RE~J1034+396 and Ark~564.
The double blackbody mimics a multitemperature optically thick disc model.
If this is the case, the fitted range of temperatures would imply
small black hole masses.
A broken power law model is preferred for the radio--loud quasar
PKS~0558$-$504 and for Ton~S~180. In the former case the X--ray emission
mechanism is probably different and likely to be related to the radio
properties. In the latter, the overall energy distribution peaks in the
far ultraviolet \cite{C98} suggesting a lower accretion disc temperature
and thus a negligible contribution in the X--ray band.

\begin{table*}
\centering
\caption{LECS$+$MECS spectral fits in the 0.1--10 keV energy range}
\vspace{0.05in}
\begin{tabular}{lcccccc}
\hline
 \multicolumn{7}{c}{Double Blackbody plus power law} \\
\hline
Source & $N^a_{H gal}$ & $kT^b_s$ & $kT^b_h$ & $\alpha_h$ & $L^c_{soft}$ & 
 SXD$^d$ \\
\hline
RE~J1034+396 & 1.5 & 55$\pm$10 & 155$\pm$25 & 1.19$\pm$0.24 & 1.6 & 23  \\
\hline
Ark~564 & 6.4 & 35$^{+17}_{-11}$ & 154$\pm$7 & 1.41$\pm$0.04 & 4 & 10 \\
\hline
 \multicolumn{7}{c}{Blackbody plus Power Law}  \\
Source & $N_{H gal}$ & $kT^b$ & $\alpha_s$/$E_{break}$ & 
$\alpha_h$ & $L^c_{soft}$ & SXD$^d$ \\
\hline 
IRAS~13224$-$3809 & 4.8 & 117$\pm$17 & ... & 0.64$\pm$0.35 & 0.9 & 9 \\
\hline
PG~1115+407$^e$ & 1.7 &  57$^{+31}_{-40}$ & 2.6/0.4 & 1.28$\pm$0.13 & 5  & 
 5 \\
\hline
IRAS~13349+2438$^e$ & 1.1  & 81$^{+7}_{-10}$ & 1.8/1.0 & 0.82$\pm$0.06 & 5 &
 2.5  \\
\hline
 \multicolumn{7}{c}{Broken Power Law}  \\
Source & $N_{H gal}$ & $\alpha_s$ & $E_{break}$ & $\alpha_h$ & 
$L^c_{soft}$ & SXD$^d$ \\
\hline 
Ton~S~180 & 1.5 & 1.68$\pm$0.07 & 2.5$^{+0.5}_{-0.9}$ & 
1.29$^{+0.12}_{-0.19}$  &  4.9 & 7 \\
\hline
PKS~0558$-$504 & 4.9 & 1.29$^{+0.16}_{-0.09}$  & 1.1$^{+0.3}_{-0.6}$  & 
1.04$\pm$0.03 & 29  & 2 \\
\hline
\end{tabular}
\par\noindent
$^a$ Galactic column density (units of 10$^{20}$ cm$^{-2}$)
\par\noindent
$^b$ Blackbody temperature (in eV)  
\par\noindent
$^c$ Unabsorbed 0.1--2 keV luminosity (units of 10$^{44}$ erg s$^{-1}$)
\par\noindent
$^d$ Soft X-ray dominance (defined as $L_{0.1-2~keV}/L_{2-10~keV}$)
\par\noindent
$^e$ The BB plus PL model is only slightly better than the Broken PL fits 
($\Delta \chi^2 \sim$ 2)

\end{table*}

Several features resembling absorption edges   
were found in the 1--2 keV ASCA spectra of 3 NLS1s including IRAS~13224$-$3809 
\cite{L97}, while line--like excess emission was
discovered around 1 keV in a few other objects observed 
with ASCA \cite{F98}, \cite{L992}, \cite{V992} including 
Ton~S~180 \cite{TGN98} and Ark~564 \cite{V991}, \cite{TGN99}.
The nature of these features is still not well understood. 
If the absorption edges are due to oxygen (the primary source of opacity in 
photoionized gas), the observed energies would imply relativistic outflows 
(see \cite{L97} for more details). Alternatively, a blend of resonance 
absorption lines in a highly ionized gas \cite{NFM99} could provide a viable
explanation. For some values of the ionization parameter, the accretion
disc reflection spectrum is rich in emission lines \cite{RFY99}.
Indeed, it has been proposed that the 1 keV emission features 
in Ton~S~180 and Ark~564 are due to a blend of such lines  \cite{TGN98} 
and/or to {\sf O VIII} recombination continuum emission \cite{V991}.

There is no convincing evidence of emission lines and/or absorption edges
in any of the 0.1--2 keV BeppoSAX spectra when the best fit models of table 
3 are adopted (e.g. Fig.~4). 
The origin of the discrepancy with the ASCA results is likely to be due
both to the different energy resolution and spectral coverage 
of the detectors onboard ASCA and BeppoSAX.
 
\begin{figure}[htb]
\centerline{\psfig{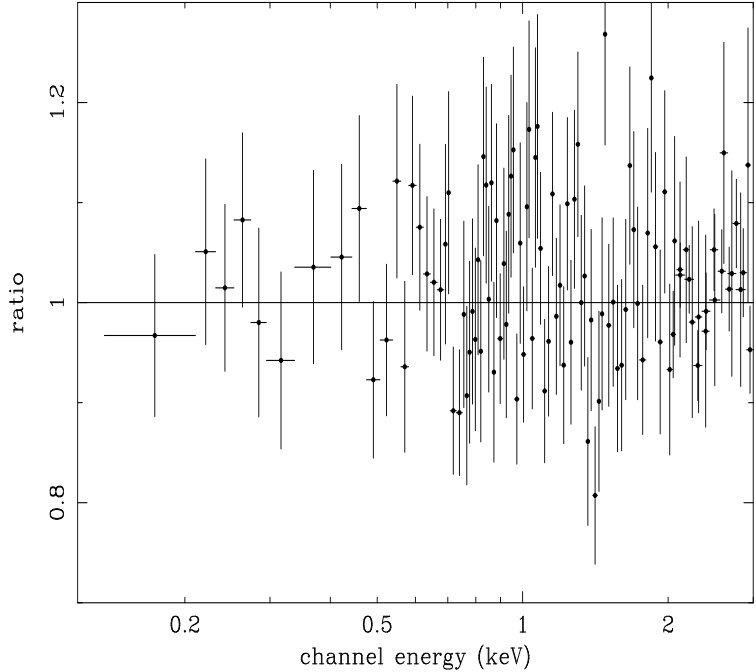}}
\caption{The residuals of the best fit model (Table 3) to the 0.1--3 keV  
spectrum of Ark~564.}
\end{figure}

In order to check whether the relatively weak emission line 
detected by ASCA could have been missed by the BeppoSAX LECS,
a simulated 100 ksec BeppoSAX observation of Ark~564 
was produced with the best fit ASCA parameters quoted by \cite{TGN99}.
Given that the simulated spectrum is well described by the 
best fit model of Table 3, it turns out that weak soft X--ray emission features 
can hardly be studied with BeppoSAX.
On the other hand, the lack of sensitivity of the ASCA instruments below 
$\sim$ 0.6 keV coupled with the calibration uncertainties at these energies,
prevent a detailed modeling of the soft X--ray continuum and 
of the strength of any emission/absorption features.
An emission line at $\sim$ 1 keV is still statistically required 
if the ASCA spectrum is fitted with a blackbody plus power law model; 
however the derived line EW is about a factor of 2 lower than that
obtained without a blackbody component. 

\section{Conclusions}

The present results fit fairly well with the hypothesis 
of a higher accretion rate relative to the Eddington rate in NLS1s
with respect to BLS1s. If the energy conversion efficiency is the
same, NLS1s should have smaller black hole masses and a correspondingly
higher accretion disc temperature. 
The steep soft excess and the good fit obtained with thermal
components are explained by a shift of the accretion disc spectrum 
in the soft X--ray band, while the rapid variability 
is naturally accounted for by the small black hole mass.
The strong soft excess could lead to a strong Compton cooling 
of the hot corona electrons and thus to a steep hard tail.
In some models the disc surface layers become strongly ionized
when the accretion rate approaches the Eddington limit, which 
fits nicely with the detection of ionized iron K$\alpha$ lines
in a few objects. Finally the optical line width is inversely proportional 
to $L/L_{Edd}$ if the broad line region is virialized 
and its radius is a function of luminosity alone \cite{LF97}, \cite{BBF96}.

The observation of ionized lines implies that reprocessing is occurring
at some level; however, the strong soft component cannot be due 
to disc reprocessing alone unless the primary emission is 
highly anisotropic or the high energy spectrum extends up to the MeV region
without any cut--off. Unfortunately the BeppoSAX sensitivity at high 
energies is not good enough to measure the shape of the high energy 
spectrum especially for steep spectrum sources, and sensitive X--ray 
observations of NLS1s at E$>$ 10 keV are not foreseen in the near future.
A reliable estimate of the amount of reprocessed radiation in NLS1s
would be also extremely important to better understand the nature of the 
soft X--ray features detected by ASCA. 

Broad band observations in the 0.1--10 keV range with better sensitivity 
and higher energy resolution will be/are being  carried out by 
{\it Chandra} and {\it XMM--Newton}.
The X--ray continuum shape and the intensity of any emission/absorption 
features will be measured with unprecedented detail. 
In addition, variability studies of the various
spectral components will be valuable to test the leading hypothesis     
of an extreme accretion rate in NLS1s.

\begin{ack}

I thank all the people who, at all levels, have made possible the SAX mission.
This research has made use of SAXDAS linearized and cleaned event
files (Rev.2.0) produced at the BeppoSAX Science Data Center.
It is a pleasure to thank all the scientists involved 
in the BeppoSAX Core Program NLS1 observations for the fruitful 
collaboration.
Partial support from the Italian Space Agency under
the contract ASI--ARS--98--119, and the Italian Ministry for University
and Research (MURST) under grant Cofin--98--02--32
are acknowledged.

\end{ack}



\end{document}